\documentclass{article}
\usepackage{graphicx}
\usepackage{listings}
\usepackage{url}
\usepackage{tabularx}

\newcommand{\footremember}[2]{%
    \footnote{#2}
    \newcounter{#1}
    \setcounter{#1}{\value{footnote}}%
}
\newcommand{\footrecall}[1]{%
    \footnotemark[\value{#1}]%
} 

\title{CF4J: Collaborative Filtering for Java}

\author{
    Fernando Ortega \footremember{u-tad}{U-tad: Centro Universitario de Tecnología y Arte Digital, Madrid, Spain}
    \and Bo Zhu \footremember{upm}{Universidad Politécnica de Madrid, Madrid, Spain} \footremember{bit}{Beijing Institute of Technology, Beijing, China} 
    \and Jesús Bobadilla \footrecall{upm} 
    \and Antonio Hernando \footrecall{upm} 
}

\date{}

\begin{document}

\maketitle

\begin{abstract}
Recommender Systems (RS) provide a relevant tool to mitigate the information overload problem. A large number of researchers have published hundreds of papers to improve different RS features. It is advisable to use RS frameworks that simplify RS researchers: a) to design and implement recommendations methods and, b) to speed up the execution time of the experiments. In this paper, we present CF4J, a Java library designed to carry out Collaborative Filtering based RS research experiments. CF4J has been designed from researchers to researchers. It allows: a) RS datasets reading, b) full and easy access to data and intermediate or final results, c) to extend their main functionalities, d) to concurrently execute the implemented methods, and e) to provide a thorough evaluation for the implementations by quality measures. In summary, CF4J serves as a library specifically designed for the research trial and error process.
\end{abstract}

\section{Introduction}

Recommender Systems (RS) provide a relevant tool to mitigate information overload problem \cite{adomavicius2005toward,adomavicius2010context}. RS act as a filter that allows to pass the relevant information to the user and blocks the irrelevant one. RS have been used to recommend a wide variety of items \cite{ricci2010introduction}: movies, books, e-commerce, educational resources, etc.
Collaborative Filtering (CF) is the most popular implementation of RS \cite{bobadilla2013recommender,herlocker1999algorithmic}. In CF recommendations are computed based on the ratings that the community of users have made over a set of items. CF can be sub-classified into \cite{pazzani1999framework,su2009survey}: 

\begin{enumerate}
    \item Memory based CF: recommendations are computed with methods that act directly on the rating matrix. Memory based methods usually use similarity metrics to obtain the distance \cite{sarwar2001item} between two users, or between two items, based on their corresponding ratings.
    \item Model based CF: recommendations are computed through a model generated from the rating matrix. Matrix factorization models \cite{hernando2016non,koren2009matrix} currently are the most popular implementations.
\end{enumerate}

To provide accurate recommendations is one of the major challenges of this big data field. A large number of researchers around the world have published papers improving different RS features such as \cite{herlocker2004evaluating}: accuracy, precision, novelty or reliability. These papers usually compare their proposed methods against different baselines, in order to highlight their improvement.  The experimentation phase requires to implement the new proposal as well as the baselines, to execute the comparison using various datasets. Any tool that simplifies the codification of the recommendation methods and/or speeds up the execution time of the experiments is desirable for RS research community.

\section{Problems and Background}

CF libraries facilitate the deployment of RS speeding up the implementation time and reducing the cost of maintenance. Several CF libraries have been released to perform this task, but the most popular ones are LibRec \cite{librec} and Apache Mahout \cite{apacheMahout}. 

LibRec is a GPL-licensed Java library, aiming to solve two classic tasks in RS, i.e., rating prediction and item ranking. This library has been developed focusing in the following features: (a) rich algorithms: LibRec has integrated more than 70 types of recommendations algorithms, (b) modularity: to optimize its maintenance the library has been divided into data processing, algorithm training and post-processing modules, and (c) efficient execution (performance): LibRec has optimized the framework and algorithm implementations as much as possible.

The Mahout project’s goal is to build an environment for fast creations of scalable performant machine learning applications. A Mahout-based CF has been developed and its engine takes users’ preferences for items and returns estimated preferences for other items. Mahout provides a rich set of components from which you can construct a customized recommender system selecting from various algorithms. Mahout is designed to be enterprise-ready; it’s designed for performance, scalability and flexibility.

In this paper we present CF4J, a new Java library designed to carry out CF research experiments. LibRec and Mahout are great CF libraries, but they have been designed for production environments rather than for research environments. In order to evaluate the capabilities of each library, we are going to study their suitability in six different software features \cite{booch2006object}: (a) extensibility, (b) performance, (c) efficiency, (d) abstraction, (e) flexibility, and (d) scalability.

The extensibility is the capability of a software to evolve with the contributions of third-party developers. Both LibRec, Mahout and CF4J have been designed thinking in their extensibility and they are free software. LibRec includes several abstract classes such: AbstractRecommenderSimilarity, to design new similarity metrics for KNN based CF; AbstractRecommender, to implement new recommendation methods; or AbstractRecommenderEvaluator, to build your own quality measures. However, despite the existence of this abstract classes, their documentation is too poor, and you must dive into the source code to understand their behavior. Mahout library contains some interfaces to enable third-party developers to tune the recommendation algorithms: ItemSimilarity and UserSimilarity interfaces allow the definition of new similarity metrics between two users and items respectively; Recommender interface represents the recommendation process of items to users; and DataModel interface represents a repository of information about users and their associated preferences for items. The Javadoc of these interfaces are good, so it is easy to transform Mahout library to your CF requirements. CF4J allows developers to extend the library through the Partible interfaces. These interfaces let developers the capability of performing any recommendation step over the users and items sets of the CF. Furthermore, several abstract classes have been included in order to facilitate the creation of most popular CF steps: UsersSimilarities and ItemsSimilarities classes simplify the creation of new similarity metrics; FactorizationModel class includes the core of CF factorization methods; and QualityMeasure class facilitates the creation of new quality measures to get the goodness of a CF. These abstract classes and interfaces are fully documented both in the Javadoc and the online CF4J’s tutorial.

Performance measure show us how effective is a software system executing a task. In this comparison, we will evaluate performance as the facility to parallelize calculations involved in the recommendation process. LibRec and Mahout delegates the parallelization to the implementation of each recommendation algorithm. They do not provide a unique parallelization framework. Developers that extend these libraries must handle the parallelization by themselves and implements parallelism using third-party parallelization tools. However, LibRec has paid great attention to the efficiency and the built-in algorithms has been optimized as much as possible. LibRec uses Javas’s ExecutorService class to create a pool of threads based on the number of available processors. Mahout has also added parallelization to some of its recommendations methods (e.g. ParallelSGDFactorizer) using ExecutorService to handle parallelization. CF4J provides a parallelization framework specifically designed for CF. This framework supplies all required tools to perform calculations easily in a parallel way over the users and items sets. All implemented CF4J algorithms use this framework.

The efficiency is the capability of a software system of ensuring the minimum memory usage. In a CF context, we will analyze how the information about users, items and ratings are stored in computer memory. LibRec stores rating information in a sparse matrix provided by the SparseMatrix class. This matrix stores data using compressed row storage and compressed column storage to minimize memory consumption. However, some information is duplicated to speed up the rating information recovery process. Mahout has been designed to run over different distributed back-ends (Apache Spark is recommended). These distributed architectures usually replicate data (ratings) among nodes following the map-reduce paradigm. The replication maximizes the efficiency of each node and improves fault tolerance of the system. CF4J sacrifices system efficiency to ensure a proper access to the data. Each rating and its pointers to the user and item are duplicated on the User and Item classes. The memory required by the system is double, but the performance of retrieving the ratings emitted by any user to any item is higher.

The abstraction is a technique for arranging complexity of computer systems. In a CF library, the abstraction level will be measured as how well the library models a real RS, that is a set of users that rate a set of items. LibRec uses low level data structures to store CF data. It provides DataModel interface to store ratings into a BiMap (a bidirectional map that preserves the uniqueness of its values as well as that of its keys). Mahout uses Preference interface to represent ratings. The access to all the ratings is managed by the DataModel interface; it contains methods to retrieve ratings using user and item codes. On the other hand, CF4J provides a data model closer to a real-world RS. It is composed of a full class hierarchy in which User and Item classes model real users and items. These classes include several methods to simplify the access to any rating emitted or received by any user or item. 

The flexibility is the capability of a software system to fit with any unexpected scenario. To measure the degree of flexibility we are trying to ask the following two questions with each library: (a) How can I retrieve the similarity between two users? (b) How can I access to the user factors after factorization process? These questions serve as example of two common needs of CF research community. In LibRec both questions must be answered as “you can not do it”. For question (a) the similarity of each pair of users is stored on the private SymmMatrix attribute of the UserKNNRecommender class. For question (b) user and item factors are stored on a protected DenseMatrix attribute of the MatrixFactorizationRecommender. Even if we can access to these non-public attributes, we need to map the user/item to its row and column in the matrix to access the data. In Mahout the answer for question (a) is “yes”: UserSimilarity class stores similarity between each pair of users and this similarity can be retrieved throughout a public method. However, the answer for question (b) is “no”: each factorization implementation (e.g. ALSWRFactorizer class) stores user and item factors in two non-public arrays and cannot be accessed from outside. In CF4J both question can be answered as “yes, you can do it easily”. For the question (a) you have a public method in the TestUser class: it returns the similarity of each user with any user of the dataset. For the question (b), factorization methods (e.g Pmf class) include public methods to retrieve the factors of any user. 

The scalability is the ability of a software system to accommodate to a growth to infinity without losing performance. Both LibRec and CF4J have been designed to run into a standalone machine, so they can only grow vertically (increment the amount of CPU and RAM of the machine). Mahout’s library runs over distributed systems (such as Spark), so it can grow vertically and horizontally (replicate the program to different machines). 

Table \ref{tab:libs-comparison} summarizes the information described in this comparison.

\begin{table}[ht]
\footnotesize
\begin{tabularx}{\textwidth}{|l|X|X|X|}
\hline
Feature       & CF4J                                                                                                                                                 & LibRec                                                                                                                                    & Mahout                                                                                                                                    \\ \hline
Extensibility & Includes generic Partible interfaces to implement any custom CF step. It also includes several abstract classes for customize most popular CF steps. & Includes some abstract classes to personalize recommendation and evaluation steps. Poor documentation.                                    & Includes abstract classes and interfaces to fully modify the recommendation process according to CF requirements.                         \\ \hline
Performance   & Provides a parallelization framework specifically designed for CF. It simplifies the coding of parallelized algorithms.                              & Parallelization is handled by each built-in algorithm. It does not provide any facility to extends the library using parallel algorithms. & Parallelization is handled by each built-in algorithm. It does not provide any facility to extends the library using parallel algorithms. \\ \hline
Efficiency    & Ratings information is duplicated between users and items to improve library’s performance.                                                          & Data is stored in a sparse matrix to minimize memory consumption.                                                                         & The library is capable to replicate data over different distributed back-ends, maximizing the efficiency of each one.                     \\ \hline
Abstraction   & Provides a full class hierarchy that contains a high-level representation of the dataset through User and Item classes.                              & Ratings are stored in a low level BiMap data structure. Ratings must be retrieved using user and item codes.                              & DataModel interface manages rating retrieval. Ratings are modeled thorough the Preference class.                                          \\ \hline
Flexibility   & Easy access to any intermediate value generated to compute recommendation.                                                                           & Null access to any intermediate value generated to compute recommendation.                                                                & Very limited access to any intermediate value generated to compute recommendation.                                                        \\ \hline
Scalability   & Vertical scalability only                                                                                                                            & Vertical scalability only                                                                                                                 & Both horizontal and vertical scalability                                                                                                  \\ \hline
\end{tabularx}
\caption{Comparison between CF4J, LibRec and Mahout CF libraries based on different software features.}
\label{tab:libs-comparison}
\end{table}

Production and research environments usually have different (and opposite) needs. Mainly, we have the need to maximize performance and efficiency in order to reduce computational costs. However, any CF library designed for research environments must satisfy the following requirements:

\begin{itemize}
    \item \textbf{Extensibility}: new ideas must be compared with state-of-the-art baselines to highlight their improvement. A proper CF research-library must be extendable for new implementations of other researchers. The extension processes should be as simple as possible. Open source projects are recommendable to warranty the library’s evolution. 
    \item \textbf{Performance}: both model based CF and memory based CF require a high computational cost to obtain recommendations. This cost can be reduced by using parallel computing. Parallelization is usually a complex task in programming. A proper CF research-library should facilitate the parallelization of the algorithms.
    \item \textbf{Efficiency}: do not waste computer memory is desirable for any software. However, in a research environment, to make easy dataset information retrieval is more important than do not waste computer memory. It is necessary to find a tradeoff between memory consumption and library simplicity.
    \item \textbf{Abstraction}: the essential information required by CF is the user set, the item set and ratings cast from users to items. This information is usually provided to the researchers as text files with lines containing the triple <user, item, rating>. A proper CF research-library should be capable to load these files and to create a data model that represents real world interactions between users and items. 
    \item \textbf{Flexibility}: some CF publications are disruptive. These types of publications require to modify crucial components / steps of the traditional CF algorithm. A proper CF research-library must provide full access to data structures, allowing researchers to replace or modify all the components and stages of the CF methods.
    \item \textbf{Scalability}: public CF datasets contain from hundreds of thousands of ratings to hundreds of millions of ratings. Nowadays, any mid-range computer has enough computational power and memory capacity to store these datasets and to run any recommendation method in a few hours. Horizontal scalability is not required in the vast majority of CF research. 
\end{itemize}

\begin{table}[ht]
\centering
\begin{tabular}{|l|l|l|l|}
\hline
Feature       & CF4J  & LibRec & Mahout \\ \hline
Extensibility & ***** & *****  & *****  \\ \hline
Performance   & ***** & **     & **     \\ \hline
Efficiency    & *     & ****   & *****  \\ \hline
Abstraction   & ***** & *      & **     \\ \hline
Flexibility   & ***** & *      & **     \\ \hline
Scalability   & *     & *      & *****  \\ \hline
\end{tabular}
\caption{Suitability of each CF library for a research environment.}
\label{tab:libs-scores}
\end{table}

Table \ref{tab:libs-scores} shows the suitability of each library for a research environment. Each feature has been scored from one to five stars. According to this table, CF4J is the most suitable CF library for research environments. CF4J has been designed following the extensibility, performance, abstraction and flexibility majors; for it, scalability and efficiency have been sacrificed. In turn, LibRec is a great library if you need a fast RS deployment due to the wide variety of recommendation methods that provides, and Mahout is the best library if you need a high scalable RS.

\section{Software Framework}

The CF4J library and its documentation is available at its GitHub repository:  \url{https://github.com/ferortega/cf4j}

\subsection{Software Architecture}

\begin{figure}[h]
\includegraphics[width=\textwidth]{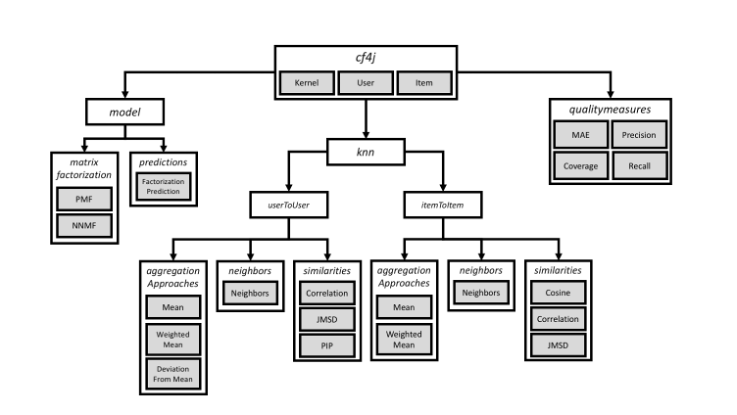}
\centering
\caption{CF4J package architecture.}
\label{fig:packages}
\end{figure}

CF4J has been designed following the packages structure shown in Figure \ref{fig:packages}. This structure has been created based on CF classification. The root of the package tree is the \texttt{cf4j} package; it contains the main classes to load, read and manipulate the CF data model. \texttt{Kernel} class loads the rating data from a text file and creates the required instances of \texttt{User} and \texttt{Item} to build de data model. The \texttt{model} package includes the implementations of model based CF methods. In this version, we have included several matrix factorization based CF implementations; other models will be easily added under this package. The \texttt{knn} package contains all the required objects to compute recommendations using memory based CF. It includes both user-to-user (\texttt{userToUser} package) and item-to-item (\texttt{itemToItem} package) approaches. Different similarity metrics (Pearson correlation, cosine, MSD, etc.), aggregation approaches (mean, weighted mean and deviation from mean) and neighborhoods creation algorithms have been added to the similarities, \texttt{aggregationApproaches} and \texttt{neighbors} sub-packages respectively. Finally, the \texttt{qualityMeasures} package contains different measures to check the goodness of the predictions (MAE and Coverage) and the recommendations (Precision and Recall).

\subsection{Software Functionalities }

CF4J has been designed to cover the CF research community needs. The main functionalities provided with the library are:

\begin{itemize}
    \item \textit{Data loading from third party text file based datasets}. CF4J reads sparse ratings matrices from text files and it creates a full class hierarchy to store this information. CF4J also divides users and items into training and tests user sets and training and test item sets.
    \item \textit{In memory data storage}. Information about users, items and ratings is stored in memory using efficient data structures that make it easy to retrieve any dataset information. Several statistical data aggregations (rating average, rating standard deviation, etc.) are also precomputed to speed up future computations.
    \item \textit{Extensible pipeline}. CF algorithms are usually composed of several sequential steps. Each of these steps generates an output that is used for the following step. CF4J includes several maps to store objects generated for any step that can be retrieved by the following step.
    \item \textit{Fast execution}. CF4J includes the Processor class to simplify the parallelizable calculations. Implementing Partible interface, CF4J concurrently executes any process over users, test users, items or test items arrays.
    \item \textit{Model-based methods support}, implementing matrix factorization machine learning solutions.
    \item \textit{Memory-based methods support}, implementing the K Nearest Neighbors approach, both in its user to user and item to item variations.
    \item \textit{Quality measures support}: both prediction and recommendation ones.
\end{itemize}

\section{Implementation and Empirical Results}

CF4J is a Java library designed to carry out experiments. Its installation is available from most popular Java repositories: Maven, Gradle and SBT. A jar file has also been packetized to be used in projects that do not use these repositories. Once the library is imported to the Java project, all classes and methods described in Section 3 are available to be used.

Experiments are easily encoded using Kernel and Processor classes. Kernel class manages all the information related with the RS. This class implements the singleton design pattern and it contains methods to load and manipulate ratings. This class also splits the users and items sets into test and training sets. Processor class handles the parallel execution of Partible implementations. A Partible is an interface that transform any CF algorithm steps to be parallelized. Parallelization is carried out over the users, test users, items or test items sets. Most of the provided example implementations have been encoded using the Partible interfaces.

Once a novel proposed research method has been implemented using CF4J, the typical experiment steps are:

\begin{enumerate}
    \item Load ratings from a text file.
    \item For each tested method:
    \begin{enumerate}
        \item Compute predictions and recommendations.
        \item Check the quality of the predictions and recommendations using quality measures.
    \end{enumerate}
    \item Show results and analyze them.
\end{enumerate}

\begin{figure}[h]
\includegraphics[width=\textwidth]{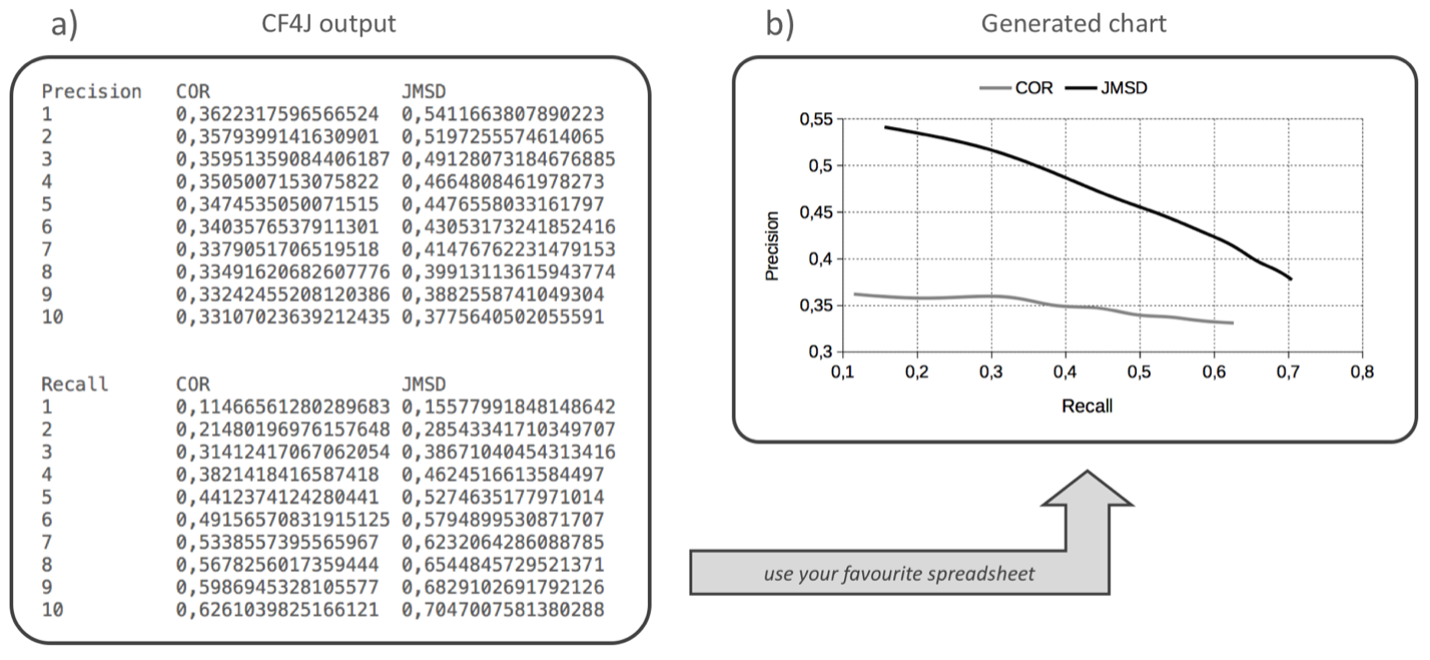}
\centering
\caption{Output of CF4J and transformation to a chart using LibreOffice spreadsheets.}
\label{fig:generate-charts}
\end{figure}

Using PrintableQualityMeasure class you can store and print results for different CF based implementations comparison. The output of the script is exportable to any spreadsheet, and it is possible to obtain different charts summarizing results. Figure \ref{fig:generate-charts} contains: a) the output of a script comparing two similarity metrics, and b) the generated chart using a spreadsheet processor.

\section{Illustrative Examples}

In this section we are going to explain, step by step, a representative example of CF experimentation using CF4J. In this example we compare MAE’s quality measure for different number of neighbors, using Pearson Correlation and JMSD as similarity metrics. Listing 1 contains the full code of the example. 

Lines 01-10 will set up the environment to perform the experiment. Lines 01-05 load ratings from a text file and define parameters to split dataset into training and test sets. In this example we are using MovieLens dataset, but any other ratings based dataset can be used. Line 07 defines the similarity metrics (COR and JMSD) used in this experiment. Line 08 defines the number of neighbors with which each similarity metric is going to be evaluated. Line 10 creates an instance of PrintableQualityMeasure class to store MAE results.

Lines 12-36 are the core of the experiment. Each combination of the parameters (each similarity metric for each number of neighbors) are tested to obtain its associated error. Line 17 and line 20 compute the similarity between of each test user with any user of the RS. The usage of Person Correlation or JMSD similarity is defined by the ‘sm’ variable. Line 27 establishes the neighbors of each test user. Line 30 calculates predictions, aggregating the neighbors’ ratings by using deviation from mean. Lines 33-34 obtain the error of predictions using the MAE quality measure. They also store the MAE values to be retrieved in the future.

Finally, we print results; line 39 prints, by using the standard output, the experiment results.

\begin{lstlisting}[language=Java, , caption=CF experiment encoded with CF4J., numbers=left, breaklines]
String dbPath = "datasets/movielens/ratings.dat";
double testUsers = 0.20; // 20% of test users
double testItems = 0.20; // 20% of test items

Kernel.getInstance().open(dbPath, testUsers, testItems, "::");

String [] similarityMetrics = {"COR", "JMSD"};
int [] numberOfNeighbors = {50, 100, 150, 200, 250, 300, 350, 400};

PrintableQualityMeasure mae = new cf4j.utils.PrintableQualityMeasure("MAE", numberOfNeighbors, similarityMetrics);

// For each similarity metric
for (String sm : similarityMetrics) {

  // Compute similarity
  if (sm.equals("COR")) {
    Processor.getInstance().testUsersProcess(new cf4j.knn.userToUser.similarities.MetricCorrelation()); 
  }
  else if (sm.equals("JMSD")) {
    Processor.getInstance().testUsersProcess(new cf4j.knn.userToUser.similarities.MetricJMSD());
  }

  // For each value of k
  for (int k : numberOfNeighbors) {

    // Find the neighbors
    Processor.getInstance().testUsersProcess(new cf4j.knn.userToUser.neighbors.Neighbors(k));

    // Compute predictions using DFM
    Processor.getInstance().testUsersProcess(new cf4j.knn.userToUser.aggregationApproaches.DeviationFromMean());
 
    // Compute MAE
    Processor.getInstance().testUsersProcess(new cf4j.qualityMeasures.MAE());
    mae.putError(k, sm, Kernel.gi().getQualityMeasure("MAE"));
  }
}

// Print the results
mae.print();
\end{lstlisting}

This example has been added to source code at the class examples. PaperExample. Furthermore, the online documentation contains other examples of the CF4J usage. We have included examples to: a) compare MAE, Coverage, Precision, Recall and F1 for several similarity metrics in knn based CF, b) compare the MAE of user-to-user and item-to-item knn based CF, and c) compare the MAE of knn based CF vs matrix factorization based CF.

\section{Conclusions}

CF4J provides an object-oriented Java framework useful for RS CF researchers. It has been created to facilitate the experimentation of new CF based recommendation algorithms. This library has been designed for research; its flexibility allows researchers to modify any CF component or step. Additionally, multiple published methods have been included to be used as baselines. A deep comparison between CF4J and most popular CF libraries (Librec and Mahout) demonstrates the superiority of CF4J for the research trial and test processes. CF4J has been designed to simplify experiments codification. The library is capable to load sparse ratings matrix from text files into a full class hierarchy. It also split users and items into training and test sets to perform cross-validation experiments. CF4J speeds up execution by concurrently running experiments; all the library’s provided methods can execute in parallel. Finally, most popular quality measures have been included into the library to compare results of novel algorithms against baselines.

\bibliographystyle{plain} % We choose the "plain" reference style
\bibliography{refs}

\end{document}